# Giant intrinsic photoresponse in pristine graphene

Qiong Ma[1], Chun Hung Lui[2], Justin C. W. Song[3], Yuxuan Lin[4], Jian Feng Kong[1], Yuan Cao[1], Thao H. Dinh[1], Nityan L. Nair[1,5], Wenjing Fang[4], Kenji Watanabe[6], Takashi Taniguchi[6], Su-Yang Xu[1], Jing Kong[4], Tomás Palacios[4], Nuh Gedik[1], Nathaniel M. Gabor[2*], Pablo Jarillo-Herrero[1*]

*Correspondence to: nathaniel.gabor@ucr.edu; pjarillo@mit.edu

[1] Department of Physics, Massachusetts Institute of Technology, Cambridge, MA 02139, USA

[2] Department of Physics and Astronomy, University of California, Riverside, CA 92521, USA

[3] Division of Physics & Applied Physics, Nanyang Technological University, 138632, Singapore

[4] Department of Electrical Engineering and Computer Science, Massachusetts Institute of Technology, Cambridge, MA 02139, USA

[5] Department of Physics, University of California, Berkeley, California 94720, USA

[6] National Institute for Materials Science, Namiki 1-1, Tsukuba, Ibaraki 305-0044, Japan

**When the Fermi level matches the Dirac point in graphene, the reduced charge screening can dramatically enhance electron-electron (*e-e*) scattering to produce a strongly interacting Dirac liquid[1-3]. While the dominance of *e-e* scattering already leads to novel behaviors, such as electron hydrodynamic flow[4,5], further exotic phenomena have been predicted to arise specifically from the unique kinematics of *e-e* scattering in *massless* Dirac systems[6-11]. Here, we use optoelectronic probes, which are highly sensitive to the kinematics of electron scattering[12-20], to uncover a giant intrinsic photocurrent response in pristine graphene. This photocurrent emerges exclusively at the charge neutrality point and vanishes abruptly at non-zero charge densities. Moreover, it is observed at places with broken reflection symmetry, and it is selectively enhanced at free graphene edges with sharp bends. Our findings reveal that the photocurrent relaxation is strongly suppressed by a drastic change of fast photocarrier kinematics in graphene when its Fermi level matches the Dirac point. The emergence of robust photocurrents in neutral Dirac materials promises new energy-harvesting functionalities and highlights intriguing electron dynamics in the optoelectronic response of Dirac fluids.**

Graphene is a model two-dimensional Dirac material with highly tunable transport and optical properties. In particular, it exhibits multiple gate-tunable photocurrent (PC) effects, including the thermoelectric PC driven by an electron temperature gradient[14-16, 20] and the photovoltaic PC generated by electric fields[12, 13, 17]. Despite the diversity, all of these PCs share a common feature – they are prominent at high charge densities[12-17, 20] but are suppressed at the charge neutrality point (CNP). For instance, the thermoelectric PC vanishes at the CNP because the Seebeck coefficient becomes zero when the chemical potential is placed at the Dirac point[14-16, 20]; the photovoltaic effect requires strong built-in fields[12, 13] or large external bias[17], which are usually enhanced by (or inevitably involve) a shift of the Fermi level from the Dirac point. Therefore, breaking the symmetry of electron-hole occupation by a finite Fermi level usually assists the generation of PCs. In this Letter, we uncover a new PC phenomenon in graphene with opposite characteristics, i.e. it appears exclusively at the Dirac point and vanishes at high charge densities.

Our observation highlights a unique aspect of a system of strongly interacting Dirac fermions, which is realized in intrinsic graphene at zero charge density. Contrary to the conventional scenario where electron-electron (*e-e*) scattering facilitates photocarrier relaxation in graphene, the unique *e-e* scattering in charge-neutral graphene can protect the new PC observed in our experiment from being relaxed. Indeed, recent works[6-10, 21] have discussed the possibility of highly tunable carrier dynamics in gated graphene. Specifically, it has been predicted that, in the presence of a finite Fermi surface, *intra*band *e-e* scattering within the linear bands can efficiently relax the photocurrent[10, 22, 23]. By contrast, when the chemical potential is at the Dirac point, the kinematic constraints of *e-e* scattering across the Dirac point can suppress the photocurrent relaxation[9-11], potentially leading to a robust photocurrent that propagates over a long distance. However, direct experimental evidence of such a novel photocurrent is still lacking. One major challenge is that the high crystalline symmetry of graphene prevents the generation of a net current, whereas near metallic electrodes or p-n junctions (where symmetries are reduced) the signals are usually overwhelmed by conventional thermoelectric and photovoltaic PCs.

Here we overcome these difficulties by fabricating graphene devices with special geometrical patterns. With such unique device geometries, we observe an anomalous PC that emerges only when the chemical potential is placed at the Dirac point. The PC appears at free graphene edges and is enhanced at edges with sharp bends. The new PC exhibits a distinct gate dependence and spatial pattern from those of conventional PCs[12-16, 20].

In our experiment, we excite graphene with a focused 850-nm continuous-wave laser and collect the PC through the source and drain electrodes at zero bias[15, 20]. PC images were obtained by scanning the laser spot across the sample (Figure 1a). All the PC data presented were taken at $T = 90$ K. In the main text, we have chosen four representative devices with different geometries (Figure 1b,f,j, 2a).

The most distinctive feature of the new PC is its unique gate dependence: it emerges exclusively at the CNP where the Fermi level matches the Dirac point. At high charge densities, we only observe conventional PC near the contacts (Figure 1c,g,k), as extensively reported in the literature[12-16, 20, 24]. As we tune graphene to the CNP, the PC at the contacts is suppressed. Instead, a pronounced PC is generated in areas far away from the contacts (Figure 1d,h,l). Such anomalous PC at the CNP, with intensity comparable to that of the contacts PC at high density, is hitherto unknown in graphene. We have observed the anomalous PC in more than ten devices with either as-exfoliated edges or plasma-etched edges on different (SiO$_2$ or BN) substrates with varying dielectric thicknesses (See supplementary information (SI)). In addition, it persists from $T = 4$ to 300 K, even after many thermal annealing cycles. This universality and robustness strongly suggest that the new PC arises from the intrinsic graphene properties rather than extrinsic factors, such as impurities or contamination[24, 25].

To further investigate the charge density dependence of the anomalous PC, we measured Device 4 with a multiple cross geometry (Figure 2). At the CNP, we observe the anomalous PC at all crossings away from the contacts (Figure 2b). When the chemical potential is tuned away from the Dirac point, the anomalous PC signals vanish abruptly at all the crossings, with nearly the same gating dependence, whereas the contact signal becomes significant at non-zero charge density (Figure 2d). We have compared the gate-dependent resistance curves and the PC curves of two devices (Device 4 and an extra device) with different graphene quality (Figure 2e). For both devices, the PC peak is narrower than the resistance peak, and the PC peak becomes sharper when the device quality improves (bottom panel in Figure 2e). This indicates that the anomalous PC emerges only at the Dirac point.

In addition to the unusual density dependence, the anomalous PC has distinct spatial patterns from those of conventional PCs. First, it is prominent at areas with geometric variation, and its magnitude and polarity depend on the local graphene geometry, as clearly shown in the comparative studies of Device 1 and Device 2 (Figure 1d and h). Second, the PC is generated only from the graphene edges rather than the bulk. In Device 3 with three rectangular graphene regions of increasing widths (Figure 1j-l), the PC appears at the neck and corner areas (Figure 1m). As the PC is spatially resolved in the lower (and wider) graphene region, we confirm that only the graphene edges contribute to the PC (white arrows). Such edge PC is different from the reported photo-Nernst current from graphene edges, which requires the application of a magnetic field[26, 27]. Third, the anomalous PC exhibits a long-range response. In Device 4, the four cross-shaped graphene sections have the same geometry but different distance from the collection contacts (Figure 2a-b). The PC intensity at the CNP does not depend on the distance between the excitation spot and the collection contacts. Strong PC appears even at >10 μm away from the contacts, a distance far exceeding the typical diffusion length scale of graphene photocarriers (<1 μm)[18, 28].

The distinctive gate dependence and spatial patterns of the observed PC are intriguing. Below we will first analyze the spatial patterns to understand the collection mechanism of the PCs, and then further explain the unique gate dependence. The PC collection (an electrically measurable PC signal) requires two steps. First, a local and *directional* PC ($j_{local}$) must be generated at the laser excitation spot. Second, this local PC must induce a global current to reach the current-collecting contacts, which are typically micrometers away from the laser excitation area.

The generation of the local PC ($j_{local}$) is possible only when the spatial circular symmetry is broken. If we photoexcite an area in bulk graphene, the PC flows out radially and cancels out (Figure 3a). But if we photoexcite the edge with broken reflection symmetry, a finite local PC can flow in the perpendicular direction (Figure 3b). However, since the photoexcitation generates electron-hole pairs, the electron and hole currents tend to cancel each other. A finite PC is possible only when these two currents are not the same. This is possible because of several reasons. First, the electrons and holes in graphene have different dispersions and group velocity at high energies[1]. Second, edge states, impurities, fringe field and strain in graphene usually induce different mobilities for electrons and holes[25, 29-33]. Based on our extended measurements (see SI) of different devices, the electron-hole current imbalance is more likely due to the first.

Once a directional $\mathbf{j}_{\text{local}}$ is generated at graphene edges, the next question is whether $\mathbf{j}_{\text{local}}$ can induce a global current to reach the contacts. As graphene is a conductor, $\mathbf{j}_{\text{local}}$ can act as a local electromotive force (emf) that drives ambient carriers. This creates a global diffusion current that flows to the collection electrodes[26, 34]. The measured PC, $I$, can be obtained by the relation $I \propto \int \mathbf{j}_{\text{local}}(\mathbf{r}) \cdot \nabla \psi(\mathbf{r}) d^2 \mathbf{r}$ according to the Shockley-Ramo theorem[34], where $\nabla \psi$ is the gradient of the electrical potential (see SI section S1). Based on this formula, $I$ is finite only when $\mathbf{j}_{\text{local}}(\mathbf{r})$ is not orthogonal to $\nabla \psi$. As discussed above, $\mathbf{j}_{\text{local}}(\mathbf{r})$ flows perpendicular to the edge. For a straight edge, $\nabla \psi$ is expected to be parallel to the edge (i.e. perpendicular to $\mathbf{j}_{\text{local}}$), hence giving rise to no global PC. However, at regions with geometrical variation (corners, necks, etc.), $\nabla \psi$ may strongly bend and lead to a nonzero global PC. These arguments naturally explain why the CNP PC is most prominent at the corner and neck regions.

To quantitatively understand the observed spatial patterns, we have calculated the $\nabla \psi$ field in various device geometries (see SI section S1). Figure 3c displays the $\nabla \psi$ field in a device with two different rectangular graphene sections of uniform conductivity. The field lines are approximately parallel to the edges within each section, with nearly zero $\mathbf{j}_{\text{local}}(\mathbf{r}) \cdot \nabla \psi(\mathbf{r})$. At the corners, however, the field lines are strongly distorted to give large $\mathbf{j}_{\text{local}}(\mathbf{r}) \cdot \nabla \psi(\mathbf{r})$ and hence strong PC (dashed circles in Figure 3c). The PC polarity also depends on the local edge geometry, as observed in our experiment. Using this scheme, we have simulated the PC images of Devices 1-4 (Figures 1e,i,m and 2c) and reproduced all major spatial PC patterns. Moreover, in this scheme, as we note, thermal currents from the edges are strongly suppressed (see SI section S8). This allows us to isolate the CNP PC from conventional signals. The excellent agreement strongly supports that a perpendicular local PC emerges at the graphene edge.

We remark that Devices 2 and 3 also exhibit noticeable PC signals along the straight edges, which cannot be reproduced by our simple model (red arrows in Figure 1h, 1l). These additional PC signals can be simulated by assuming a different conductivity at the edge from the bulk, causing bending of field lines even near straight edges, which is a possibility given recent work on edge states in graphene[29, 31, 32] (see SI section S2.4 for more discussion).

Now we turn to explain the unique gate dependence, i.e., why the PC emerges only when the chemical potential is shifted to the Dirac point, which indicates unusual charge carrier dynamics at the CNP. Based on the above discussion, although the generation of $\mathbf{j}_{\text{local}}$ is allowed at regions with broken reflection symmetries, the collection of a global PC requires $\mathbf{j}_{\text{local}}$ to propagate a considerable distance from the edge so as to couple to the distorted field lines in the Shockley-Ramo-type model (SI section S1.3). Rapid scattering with other electrons, phonons and impurities are usually considered as strongly suppressing $\mathbf{j}_{\text{local}}$[23]. Recent theory, however, predicted robust local PC propagation in charge-neutral graphene when *e-e* scattering dominates[10].

To explore the electron dynamics in graphene with strong *e-e* interactions, we first consider the regime of finite chemical potential, in which *intra*band processes dominate *e-e* scattering. As an illustration, we consider an excited electron (hole) with initial momentum $\hbar\vec{k}_1$ and energy $\hbar v k_1$ and its subsequent scattering with a second electron at $\vec{k}_2$ in the Fermi sea. The energy and momentum conservation and Pauli exclusion require:

$$k_1 + k_2 = k_1' + k_2' \qquad (1)$$

$$\vec{k}_1 + \vec{k}_2 = \vec{k}_1' + \vec{k}_2' \qquad (2)$$

$$k_1, k_1', k_2' > k_F > k_2 \qquad (3)$$

The left and right sides of the equations correspond to the states before and after the scattering (Figure 4a). These relations constrain the final wave vectors ($\vec{k}_1'$ and $\vec{k}_2'$) in an ellipse set by the initial wave vectors ($\vec{k}_1$ and $\vec{k}_2$), with differing directions (Figure 4b). Because the electrons travel at constant speed along the wave vector in the Dirac cone, the total current is not conserved after the scattering. Similarly, the current also changes for an excited hole scattered with an electron within the Fermi sea (Figure 4c). In this case, the different energy conservation relation $k_1 - k_2 = k_1' - k_2'$ defines a hyperbolic constraint for the final wave vectors (Figure 4d). In a quantitative calculation[10], the total current of an excited electron-hole pair can be shown to decrease after scattering with other electrons. As the *e-e* scattering time for typical doping levels is extremely short in graphene (<10 fs)[9, 18, 35, 36], the local PC is expected to be quenched abruptly and will thus not contribute to the global PC.

The PC relaxation is, however, very different when the chemical potential is at the Dirac point. In this special case, *intra*band processes described above are strongly suppressed due to the vanishing Fermi surface. The only remaining *e-e* scattering process to relax the excited electron (hole) is to produce another electron-hole pair at lower energies (Figure 4e)[9, 10, 21]. This *inter*band process requires the momenta of the initial and final states to stay along the same line due to the energy and momentum conservation (Figure 4f and SI section S6)[9, 21]. Such a collinear scattering preserves the total current[10]. Therefore, the PC does not decrease even after many *e-e* scattering events. The local PC can then travel far away from the edge and contribute to the global PC. This picture accounts excellently for our observed edge PC, which appears exclusively at the CNP but vanishes at finite chemical potential. Such sensitive gate tunability of the PC cannot be found in conventional materials with parabolic dispersion, where *e-e* scattering always preserves the current[10] (see SI section S5-S7 for more discussions). We want to point out that although these collinear processes are in principle suppressed due to the vanishing phase space, creating a bottleneck for electron-hole pair production in the photoexcitation cascade, recent theory predicts that many-body effects can unblock such constraints and lead to sharp near-collinear angular distribution of secondary carriers[11].

Our observation of highly tunable CNP PC not only provides new principles for the design, implementation and optimization of energy-harvesting and ultra-efficient photodetector devices based on graphene, but also points out several remarkable characteristics of this prototypical 2D Dirac material. First, the carrier relaxation is dominated by *e-e* scattering under our experimental conditions. Since the electron-phonon and electron-impurity scattering processes depend weakly on the Fermi energy, the sharply emergent Dirac-point PC suggests that the contribution from these current-relaxation processes is significantly weaker compared to that of the *e-e* scattering[9, 18, 21, 35, 37, 38]. Second, our results demonstrate that the relaxation of highly non-equilibrium carriers can be strongly affected by the low-energy Dirac cone of graphene, giving rise to observable consequences even in steady-state measurements. Such a remote yet strong coupling between high-energy carriers and low-energy band properties provides a novel approach to probe the low-energy physics of Dirac materials with optical and optoelectronic means.

Finally, our results give tantalizing evidence toward probing optoelectronic response of Dirac electron fluids. After measuring more than 10 graphene devices at $T = 4 – 300$ K, we observed strong CNP PC only in devices with high mobility ($\mu > 10,000$ cm$^2$/(V·s)). In addition, the Dirac point PC is strongly enhanced at intermediate temperatures (60 -120 K) and suppressed outside that temperature range (Figure S10, SI). Such dependence on sample quality, charge density and temperature are consistent with the characteristics of hydrodynamic electron flow in graphene[4]. Indeed, our experiment approaches the conditions of the Dirac fluid and suggests possible intriguing optoelectronic dynamics in Dirac fluids, which is a fascinating new research area that has been relatively unexplored so far.

## Methods

**Device fabrication.** We obtained graphene samples by mechanical exfoliation of graphite crystals. The graphene samples for Devices 1 – 4 were deposited on degenerately doped silicon substrates with a 285 nm oxide epilayer. We patterned graphene into different geometries by e-beam lithography and O$_2$ plasma etching, and attached 0.8/80-nm Cr/Au electrodes by thermal evaporation. Our devices exhibit low residue doping and no photogating effect. The mobility of Devices 1-4 are between 10,000 cm$^2$/(V·s) and 20,000 cm$^2$/(V·s).

**Scanning photocurrent measurements.** We mounted the devices in a Janis ST-500 helium optical cryostat with tunable temperature down to $T = 4$ K. The electrical feed-throughs and the optical windows in the cryostat allow us to measure the source-drain current of the devices under simultaneous laser illumination. We focused the laser ($\lambda = 850$ nm) onto the samples with a spot diameter of 1 μm by using a 60X microscope objective. The laser spot was scanned over the graphene devices using a two-axis piezoelectrically controlled mirror. We recorded both the direct reflection of the laser and the photocurrent signal as a function of the laser position. Comparison of the reflection image and the photocurrent image allows us to identify the position of the photocurrent signal on the devices.

## Acknowledgements

We thank Frank Koppens, Klaas-Jan Tielrooij, Mark Lundeberg, Achim Woessner and Oles Shtanko for stimulating discussions. We also thank Yilin Sun and Bingnan Han for help with device fabrication. Work in the PJH group was partly supported by the Center for Excitonics, an Energy Frontier Research Center funded by the US Department of Energy (DOE), Office of Science, Office of Basic Energy Sciences (BES) under Award Number DESC0001088 (fabrication and measurement) and partly through AFOSR grant FA9550-16-1-0382 (data analysis), as well as the Gordon and Betty Moore Foundation's EPiQS Initiative through Grant GBMF4541 to PJH. This work made use of the MRSEC Shared Experimental Facilities at MIT, supported by the National Science Foundation under award number DMR-14-19807 and of Harvard CNS, supported by NSF ECCS under award no. 1541959. YL, TP, WF, JK, SYX and NG acknowledge funding support by the STC Center for Integrated Quantum Materials, NSF Grant No. DMR-1231319. YL and TP also acknowledge the U.S. Army Research Office through the MIT Institute for Soldier Nanotechnologies, under Award No. W911NF-18-2-0048. K.W. and T.T. acknowledge support from the Elemental Strategy Initiative conducted by the MEXT, Japan, JSPS KAKENHI Grant Numbers JP18K19136 and the CREST (JPMJCR15F3), JST.

**Author Contributions**

QM and NMG conceived the experiment; QM and NLN fabricated the devices; YL fabricated additional devices shown in the SI under the supervision of TP; QM, THD and NMG carried out the photocurrent measurements; QM, CHL and YC analyzed and simulated the data under supervision from PJH; WF and JK grew the CVD graphene; KW and TT synthesized the BN crystals; JCWS and JFK contributed to theoretical discussions, QM, CHL, JCWS, SYX, NG and PJH co-wrote the paper with input from all the authors.

**Additional Information**

Correspondence and requests for materials should be addressed to NMG and PJH.

**Competing Financial Interests**

The authors declare no competing financial interests.

**Figures and captions**

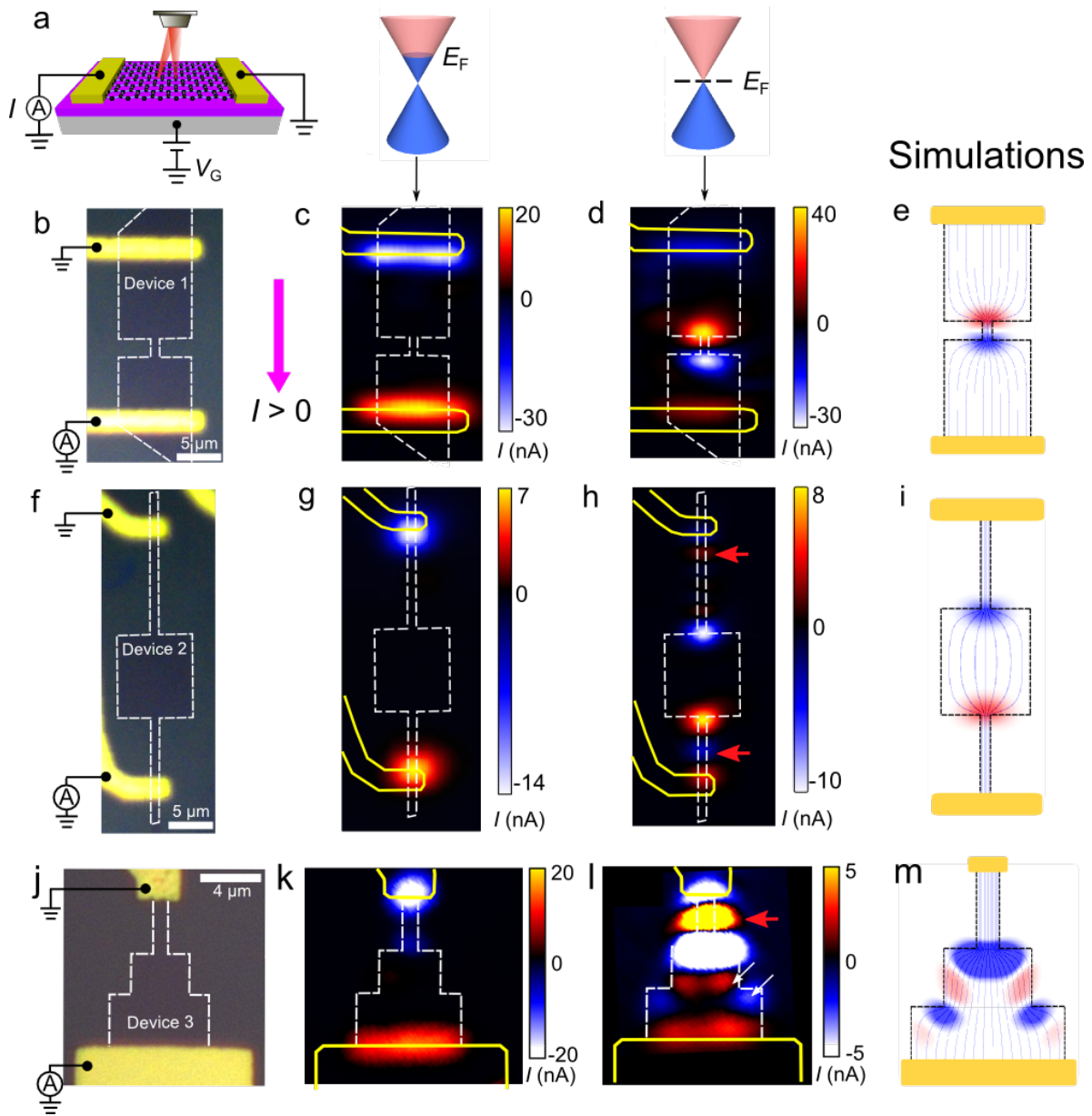

**Figure 1| Intrinsic long-range edge photocurrent (PC) in charge-neutral graphene with different geometries. a,** Schematic of a back-gated graphene device with scanning laser excitation. The photocurrent is measured in a short-circuit configuration. **b-d,** Optical image (b) and scanning PC images (c-d) of Device 1 with a narrower middle graphene channel. At high charge density (c), the PC is mainly generated at the contact areas. At the charge neutrality point (CNP) (d), however, significant PC emerges at the two ends of the middle graphene region. **e,** Simulation of the PC image for Device 1. **f-i,** Similar figures as (b-e), for Device 2 with a wider middle graphene section. **j-m,** Similar figures as (b-e), for Device 3 with three rectangular graphene regions with increasing widths. The white arrows in (l) indicate that the PC is generated from the graphene edge. The red arrows in (h) and (l) denote the PC along the straight edges, which cannot be accounted for by our simple model with uniform graphene conductivity.

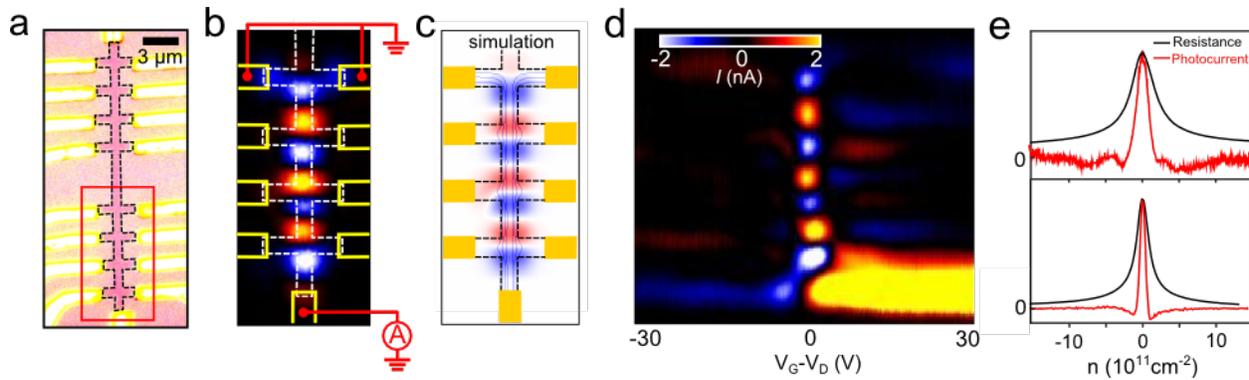

**Figure 2| Distance and gate dependence of the CNP photocurrent in graphene. a-b,** Optical image (a) and scanning PC image (b) of Device 4 with four cross-shaped graphene sections. The four sections have different distance from the collection electrodes located at the upper and lower ends of the device. **c,** Simulation of the PC image of Device 4. **d,** The PC intensity along the vertical middle channel as a function of gate voltage. **e,** Top: representative density-dependent PC in Device 4, in comparison with the resistance curve. Bottom: similar PC and resistance curves for another device with a shaper Dirac peak.

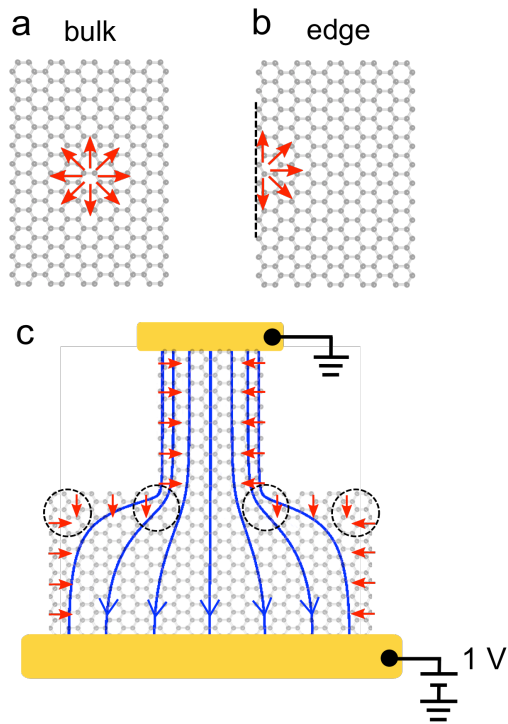

**Figure 3| Long-range edge photocurrent in graphene within a Shockley-Ramo-type scheme. a,** In the bulk area of graphene, the laser excitation induces circular PC with zero net current (red arrows). **b,** Near the edge with lower symmetry, finite net PC becomes possible. **c,** Generation of long-range PC from the graphene edges in a Shockley-Ramo-type scheme. The blue lines are the weighting field lines of the device with the source at 1V and the drain at 0V. The red arrows represent the local edge PC. The dashed circles highlight the regions with strong contribution to the total PC.

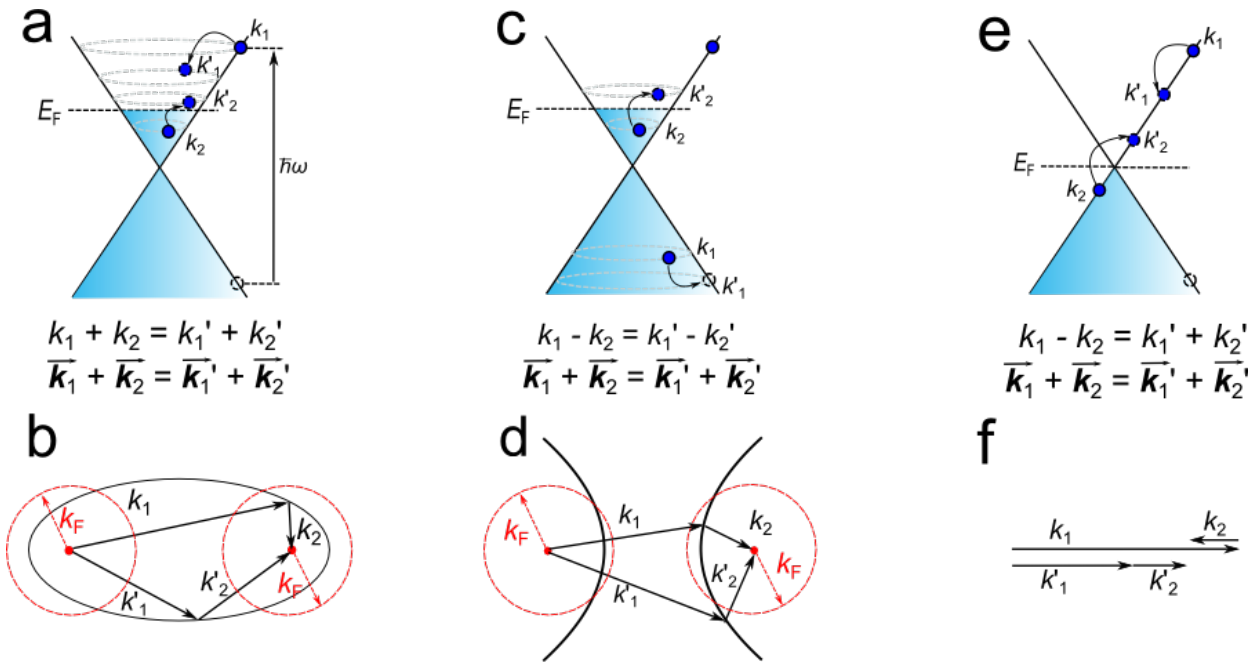

**Figure 4| Suppression of photocurrent relaxation at the charge neutrality point. a-b,** The allowed *e-e* scattering processes for an excited electron in graphene with positive chemical potential. The conservation of energy and momentum (represented by the equations) constrain the electron wave vectors to an ellipse. **c-d,** Similar scattering diagrams for an excited hole. The electron wave vectors are constrained to a hyperbola. **e-f,** Scattering diagram for charge-neutral graphene. In this case, the *e-e* scattering is limited only to collinear processes, with all four electron wave vectors in the same line. The total current is preserved after the scattering.